\documentclass[prl,aps,twocolumn,superscriptaddress,showpacs,floatfix]{revtex4-1}
\usepackage{amsmath,amssymb,multirow,epsfig,bm,dcolumn,pstricks}

\usepackage{ulem}

\renewcommand{\vec}{\bold}

\begin{document}
\title{BEC-BCS crossover and universal relations in unitary Fermi gases}

\author{S. Gandolfi}
\affiliation{ Theoretical Division,
Los Alamos National Laboratory,
Los Alamos, NM 87545}

\author{K. E. Schmidt}
\affiliation{Department of Physics, Arizona State University, Tempe,
AZ 85287, USA}

\author{J. Carlson}
\affiliation{ Theoretical Division,
Los Alamos National Laboratory,
Los Alamos, NM 87545}

\begin{abstract}
The contact parameter in unitary Fermi Gases governs the short-range correlations,
and high-momentum properties of the system.  We
perform accurate quantum Monte Carlo calculations with highly optimized
trial functions to precisely determine this parameter at T=0, demonstrate its universal
application to a variety of observables, and determine the regions
of momentum and energy over which the leading short-range behavior
is dominant. We derive Tan's expressions for the contact parameter
using just the short-range behavior of the ground-state many-body wave function,
and use this behavior to calculate the two-body distribution
function, one-body density matrix, and the momentum distribution
of unitary Fermi gases; providing a precise value of the contact
parameter that can be compared to experiments.
\end{abstract}


\maketitle 

The experimental realization and concurrent
theoretical calculations of two-component unitary Fermi gases
with short-range interactions offer a unique opportunity to test our
understanding of strongly-interacting Fermi systems, and to study
their structure and dynamics.
The low-energy properties of the system 
are governed by the Bertsch parameter $\xi$,
the pairing gap $\delta$, and have been studied extensively in the
literature~\cite{Giorgini:2008,*Carlson:2005,*Carlson:2008,Schirotzek:2008,*Lee:2008}.  
The short-range correlations of the system
are, in contrast, governed by the many-body wave function at small
interparticle separations, as encoded in the contact parameter $C$.

Tan, in a series of papers~\cite{Tan_a:2008,Tan_b:2008,Tan_c:2008},
showed that in Fermi gases, if the effective range of the interaction 
is much smaller than any other length scale of the system, 
several universal relations occur, and related them to a parameter $C$ he called
the contact parameter (see also Ref.~\cite{Braaten:2010} for a review).
In particular, the large momentum tail of the momentum distribution $N(k)$,
behaves like $C/k^4$. This same parameter gives the small distance
behavior of the two-body distribution function, and the derivative of the
ground-state energy with respect to the inverse of the scattering length.
Having multiple phenomena that depend on a single universal parameter means that
the parameter can be calculated and measured in multiple ways, and in the range of
validity of the experiments and the calculations, they must give the same results for
the parameters. We employ highly-optimized trial wave functions and
accurate quantum Monte Carlo to calculate several of these observables, and
to extract the contact parameter.  We thus produce more accurate results
for the leading behavior and simultaneously determine the regimes where
it is dominant.

Recently, several experimental measurements have measured $C$ using
a variety of techniques~\cite{Navon:2010,*Kuhnle:2010,*Stewart:2010,*Hu:2010}.
Values for the contact parameter have been calculated from previous results
from quantum Monte Carlo~\cite{Lobo:2006,*Combescot:2006,*Drut,*Su:2010,*Abe:2009} and other methods
~\cite{Palestini:2010,*Chen:2007,*Lee:2008b}.
However, previous quantum Monte Carlo
calculations give values of the contact at unitarity that disagree with
each other at the 5 to 10 percent level. Here we show that if the calculations are carefully
optimized and extrapolated to zero range, our quantum Monte Carlo results agree with
each other within statistical errors, less than 0.5 percent, giving clear numerical
evidence of Tan's predicted universal contact parameter and its behavior around
unitarity. Our results provide a benchmark prediction for
low temperature experiments.

We 
perform variational and fixed-node diffusion Monte Carlo (VMC and DMC)
calculations of a system of a homogeneous system of fermions interacting
with a short range potential. Fixed-node diffusion Monte Carlo results
produce upper bounds to the ground-state energy of the system depending only
upon the nodes (zeroes) of the trial wave function. We carefully optimize the trial wave
functions, and obtain the best upper bounds
to date for the ground state energy. Observables other than the ground-state
energy are calculated by extrapolating the variational and mixed estimates:
$O_V = \langle \Psi_T | O | \Psi_T \rangle $ and 
$O_m = \langle \Psi_0 | O | \Psi_T \rangle $ , $\langle O \rangle
\approx 2 O_m - O_V.$
After suitable optimizations the extrapolations are very small.
We calculate the contact parameter in several different ways and show they all
give results consistent with each other and with recent experiments.

Tan and later
others~\cite{Tan_a:2008,*Tan_b:2008,*Tan_c:2008,Braaten:2008,*Combescot:2009,*Werner:2009}
derived expressions for his contact
parameter using a variety of methods. These results can be understood
as coming from the behavior of the many-body wave function when
two unlike spin particles are separated by a distance $r$
small compared to the average
particle separation, $r_0$, but outside the range of the potential, $R$,
\begin{equation}
\label{eq.1}
f(r) = \frac{A(1-a^{-1} r)}{r} \,,
\end{equation}
($a$ is the two-body scattering length) which is Eq. 1 in~\cite{Tan_a:2008}
and $A^2$ will be seen to be proportional
to Tan's contact parameter $C$. The unlike spin two-body distribution
function will be given by $f^2(r)$ in this same range
\begin{equation}
\label{eq.2}
g_{\uparrow\downarrow}(r) = A^2(r^{-2}-2a^{-1}r^{-1} + \dots) \,,
\end{equation}
where $g_{\uparrow\downarrow}(r \rightarrow \infty) = \frac{1}{2}$ for
an unpolarized system.
The momentum distribution summed over both spins
will also be dominated by this
short range part of the wave function, so for $k$ much greater than the
Fermi momentum $k_F$ but much less than the inverse potential range, we have
\begin{equation}
\begin{split}
&N(k) =
n \int d^3r \, dR \,
\Psi(\vec r_1+\vec r,\dots,\vec r_N)
\Psi(\vec r_1,\dots,\vec r_N) e^{-i\vec k \cdot \vec r}\\
&= n^2 \int d^3r \, d^3r' \, \frac{f(|\vec r+ \vec r'|)}{f(r')}
g_{\uparrow\downarrow}(r')
e^{-i\vec k \cdot \vec r}\\
&
= n^2 \int d^3r \, d^3r' \, f(|\vec r+\vec r'|)f(r') e^{-i\vec k\cdot \vec r}
= \frac{16\pi^2 n^2 A^2}{k^4} \,,
\nonumber
\end{split}
\end{equation}
where $dR$ indicates integration over $r_1,\dots, r_N$,
and $n$ is the number density.
Fourier transforming the momentum distribution and the two-body distribution
functions gives the behavior for the one-body density matrix (normalized
to 1 at the origin) and the opposite
spin static structure factor (which goes to $\tfrac{1}{2}$ for
$k\rightarrow\infty$) of
\begin{eqnarray}
\rho^{(1)}(r) &=& 1 - 2\pi n A^2 r + \dots
\nonumber\\
S_{\uparrow\downarrow}(k) -\frac{1}{2} &=&
\frac{2\pi^2 nA^2}{k} \left [ 1-\frac{1}{4\pi a k} \right ] + \dots \,.
\end{eqnarray}
Tan also related the contact parameter to the derivative of the energy
with respect to the inverse scattering length. Changing the
scattering length by changing the potential with the mass fixed and
using the Hellman-Feynman theorem~\cite{Hellmann:1937,*Feynman:1939}
\begin{equation}
\frac{dE\quad}{da^{-1}} = \frac{n}{2} \int d^3r \, g_{\uparrow\downarrow}(r)
\frac{d v(r)}{da^{-1}}
\end{equation}
where $E$ is the energy per particle.
Since $v(r)$ is nonzero only inside $R$, where the two-body potential
is very strong, $g(r)$ can be replaced with $f^2(r)$ where
$\frac{\hbar^2}{m} \nabla^2 f(r) = v(r) f(r)$ and the integration
taken over a sphere of radius $R$. Therefore
\begin{equation}
\begin{split}
&\frac{dE\quad}{da^{-1}} = \frac{n}{2} \left [ \frac{d}{da^{-1}} \int d^3r \,
f^2(r) v(r) -2\int d^3r \, f(r) v(r) \frac{df(r)}{da^{-1}} \right ]
\\
&= 
\frac{\hbar^2 n}{2m} \left [ \frac{d}{da^{-1}} \int d^3r \,
f(r) \nabla^2 f(r) -2\int d^3r \,[\nabla^2 f(r)]\frac{df(r)}{da^{-1}}
\right ]
\\
&= \frac{2\pi \hbar^2 n}{m} R^2 \left .  \left [
\frac{d}{da^{-1}}
f(r) \frac{df(r)}{dr} - 2 \frac{df(r)}{da^{-1}}\frac{df(r)}{dr}
\right ] \right |_{r=R} \,.
\nonumber
\end{split}
\end{equation}
This only depends on $f(r)$ around $R$, and using Eq. \ref{eq.1}
the result is
\begin{equation}
\label{eq.7}
\frac{dE\quad}{da^{-1}} = -\frac{\hbar^2 2 \pi n A^2}{m} \, 
\rightarrow C = 8\pi^2 n^2 A^2.
\end{equation}

The equation of state and therefore Tan's $C$
\footnote{
Some authors define the contact as an extensive
quantity ${\cal C} = \Omega C$, where $\Omega$ is the volume, and report
the unitless intensive quantity $\frac{\cal C}{Nk_F} = 3\pi^2 \frac{C}{k_F^4}$.
}
are
conventionally parametrized around unitarity as~\cite{Tan_b:2008}
\begin{eqnarray}
\frac{E}{E_{FG}} &=& \xi-\frac{\zeta}{k_F a}-
\frac{5\nu}{3(k_F a)^2}+\dots \,,
\nonumber\\
\frac{C}{k_F^4}&=&\frac{2}{5\pi} \left [ \zeta +\frac{10\nu}{3(k_F a)} +\dots
\right ]
\label{eq:eos}
\end{eqnarray}
where $E_{FG} = \frac{3\hbar^2 k_F^2}{10 m}$
is the infinite system free gas energy per particle.
At unitarity we have several quantities related to $\zeta$:
\begin{eqnarray}
\rho^{(1)}(r) &\rightarrow& 1-\frac{3}{10} \zeta k_F r \,,\,\,\,\,
N(k) \rightarrow \frac{8}{10\pi} \zeta \frac{k_F^4}{k^4} 
\nonumber\\
g_{\uparrow\downarrow}(r) &\rightarrow& \frac{9\pi}{20} \zeta (k_F r)^{-2} \,,\,\,\,\,
S_{\uparrow\downarrow}(k) \rightarrow \frac{3\pi}{10} \zeta \frac{k_F}{k} \,.
\end{eqnarray}

We use Quantum Monte Carlo (QMC) techniques to accurately solve 
the many-body ground state, and compute properties of the unitary Fermi gas.
Our QMC calculations
use the many--body Hamiltonian,
\begin{equation}
H=\sum_{i=1}^N \frac{p_i^2}{2m}
-v_0 \frac{8\hbar^2}{m r_e^2} \sum_{i\uparrow,j\downarrow} 
\frac{1}{\cosh^2(2 r_{ij}/r_e)} \,,
\end{equation}
where the two--body interaction is a a short--range potential taken only
between opposite spin particles.
At unitarity, $v_0 = 1$ and the
effective range is $r_e$.
The scattering length
and effective range can be tuned by changing $v_0$ and $r_e$.
The limit of zero effective range (dilute system) is reached by 
taking $r_e \ll r_0$, with $r_0=(3/(4\pi n))^{1/3}$.
The unitary limit is approached when $r_0 \ll a$ where $a$ is the
scattering length of the two--body interaction. At unitarity the
details of the interaction are not important, and the only scale
of the system is given by its Fermi momentum $k_F$.
The ansatz for the many--body trial
wave function is the same as previously used in other QMC
calculations~\cite{Carlson:2003,*Chang:2004,*Gezerlis:2009}:
\begin{eqnarray}
 \Psi_T =  \prod_{ij} f_J(r_{ij'}) \,\, \Phi_{\rm BCS} \,,\,\,
 \Phi_{\rm BCS} = {{\cal A}}[\phi(r_{11'}) \phi(r_{22'}) ... \phi(r_{nn'})] 
 \nonumber
\end{eqnarray}
where ${\cal A}$ antisymmetrizes the like spins, and
the unprimed coordinates are for up spins and the primed are for
down spins and $n=N/2$. The pairing function is
\begin{eqnarray}
\phi({\vec r}) &=&\tilde{\beta} (r) +
\sum_{n} a(k_n^2) \exp [ i \vec k_n \cdot \vec r]~,
\nonumber\\
\tilde{\beta}(r)&=& \beta(r)+\beta(L-r)-2 \beta(L/2)~,
\nonumber\\
\beta (r) &=& [ 1 + c b  r ]\ [ 1 - \exp ( - d b r )]
\frac{\exp ( - b r )}{dbr}~.
\end{eqnarray}
The function $\tilde{\beta}(r)$
has a range of $L/2$,  the value of $c$ is chosen such that it has
zero slope at the origin.
\begin{figure}[ht]
\includegraphics[width=0.4\textwidth]{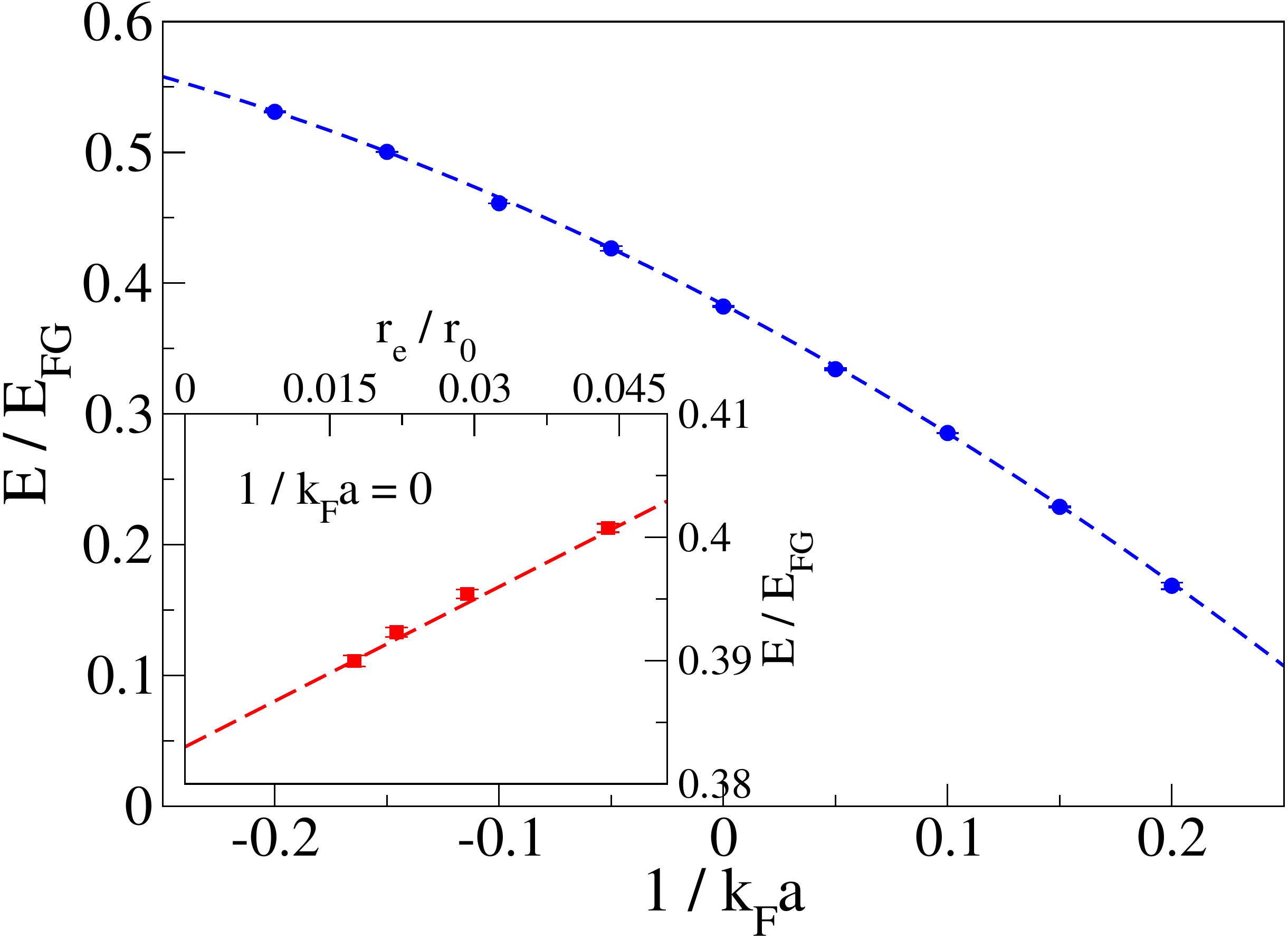}
\caption{
(color online) Energy per particle in the BEC-BCS crossover regime in units of $E_{FG}$
as a function of the scattering length $a$. The QMC points are the
results of extrapolations to $r_e \rightarrow 0$ limit. In the inset
we show the extrapolation at unitarity.
}
\label{fig:eos}
\end{figure}
\begin{figure}[ht]
\includegraphics[width=0.4\textwidth]{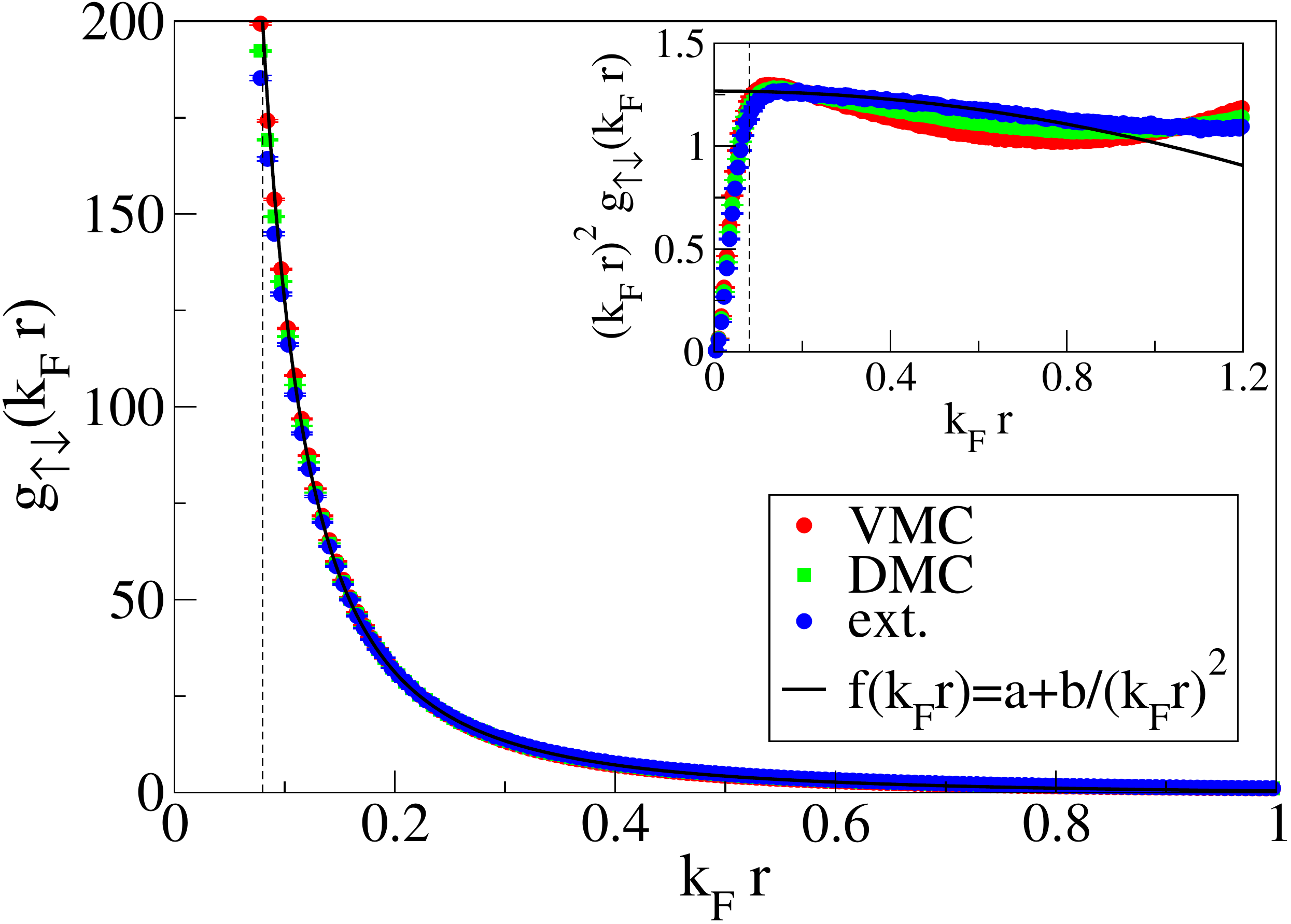}
\caption{(color online) The calculated spin-opposite two-body distribution function at small distance $r$;
the effective range of the interaction is $k_F\,r_e\approx0.08$ (vertical black dashed line).
The VMC (red), mixed (green) and extrapolated (blue) results are shown. The extrapolated QMC results
are used to fit the function giving $b=1.2678$. In the inset we show the same functions
multiplied by $(k_Fr)^2$.
}
\label{fig:gofr}
\end{figure}

The variational wave function has been carefully optimized;
in particular we optimize the pairing orbitals entering in the wave
function by using VMC to minimize the energy~\cite{Sorella:2001}. The fixed-node
DMC energies do not depend on the Jastrow function $f_J$.  Our
simulations are performed with 66 particles in a periodic box, and
we study the effect of the effective range of the interaction by
changing $r_e$ and extrapolating to the $r_e \rightarrow 0$ limit.
The results of 66 particles is very close to
the infinite limit~\cite{Forbes:2010}.
Careful optimization of the variational wave function significantly
improves the energy upper bounds.
At unitarity, the best previous QMC results
using 66 particles are $\xi=0.42(1)$ fixing $r_e/r_0\approx
0.08$~\cite{Carlson:2005}, and $\xi=0.42(1)$ using $r_e/r_0\approx
0.01$~\cite{Astrakharchik:2004}. Our new estimate is $\xi=0.4069(5)$
and $\xi=0.3923(4)$ with $r_e/r_0\approx 0.07$ and $0.02$ respectively.
The parameters for $\phi$ at unitarity are $b=0.5k_F$, $d=5$ and the nonzero
$a(k^2)$ are given in Table \ref{t1}.
\begin{table}
\begin{tabular}{|cc|cc|}
\hline
$\frac{L^2}{4\pi^2} k^2$ & $a(k^2)$ &
$\frac{L^2}{4\pi^2} k^2$ & $a(k^2)$\\
\hline
0 & 0.00198 &         5 & 0.000190  \\
1 & 0.00250 &         6 & 0.000200  \\
2 & 0.00194  &        8 & 0.000167 \\
3 & 0.00081  &        9 & 0.000163 \\
4 & 0.00033  &       10 & 0.000120 \\
\hline
\end{tabular}
\caption{The optimized
plane wave coefficients at unitarity for the pairing function.
}
\label{t1}
\end{table}
Improved optimization of the trial wave function lowers
the fixed-node energy by 4--7\%.  Careful extrapolation to $r_e\rightarrow
0$ limit is also important. We show an example at unitarity
in the inset of Fig. \ref{fig:eos} where we plot QMC points at
different effective ranges, and their extrapolation.  Using more
points and a more complex fit typically provides a somewhat lower
upper bound to the energy~\cite{Forbes:2010}; such a correction is
about 0.002 to $\xi$.

We optimized the many--body wave function for systems
with different scattering lengths and for each value of $k_F a$ we
repeated the extrapolation of $r_e$. Our results of $\xi(k_F a)$
are shown in Fig. \ref{fig:eos}.
Fitting the QMC points shown in Fig. \ref{fig:eos} gives the values
$\xi=0.383(1)$, $\zeta=0.901(2)$ and $\nu=0.49(2)$.
Using Eq. \ref{eq.7} we predict
\begin{equation}
\label{eq:c-eos}
\frac{C}{k_F^4}=\frac{2\zeta}{5\pi}=0.1147(3) \,.
\end{equation}

An alternative direct method for calculating
the contact can be obtained by computing correlation functions
at unitarity. For example, the pair distribution function is shown
in Fig. \ref{fig:gofr}, where we compare the VMC result with the
mixed estimate computed with DMC. The two results are almost identical
and differences appear only for very small distances. The value of
$\zeta$ is obtained by fitting $g_{\uparrow\downarrow}(k_F r)$
in the range $r_e\ll r\ll k_F^{-1}$ using the function $a+b/r^2$.
The fit gives $b=1.2678(1)$. Using Eqs. \ref{eq.2} and \ref{eq.7},
gives the value for
$\zeta=0.897(2)$ in good agreement with the result extracted
from Eq. \ref{eq:eos}.

\begin{figure}[ht]
\includegraphics[width=0.4\textwidth]{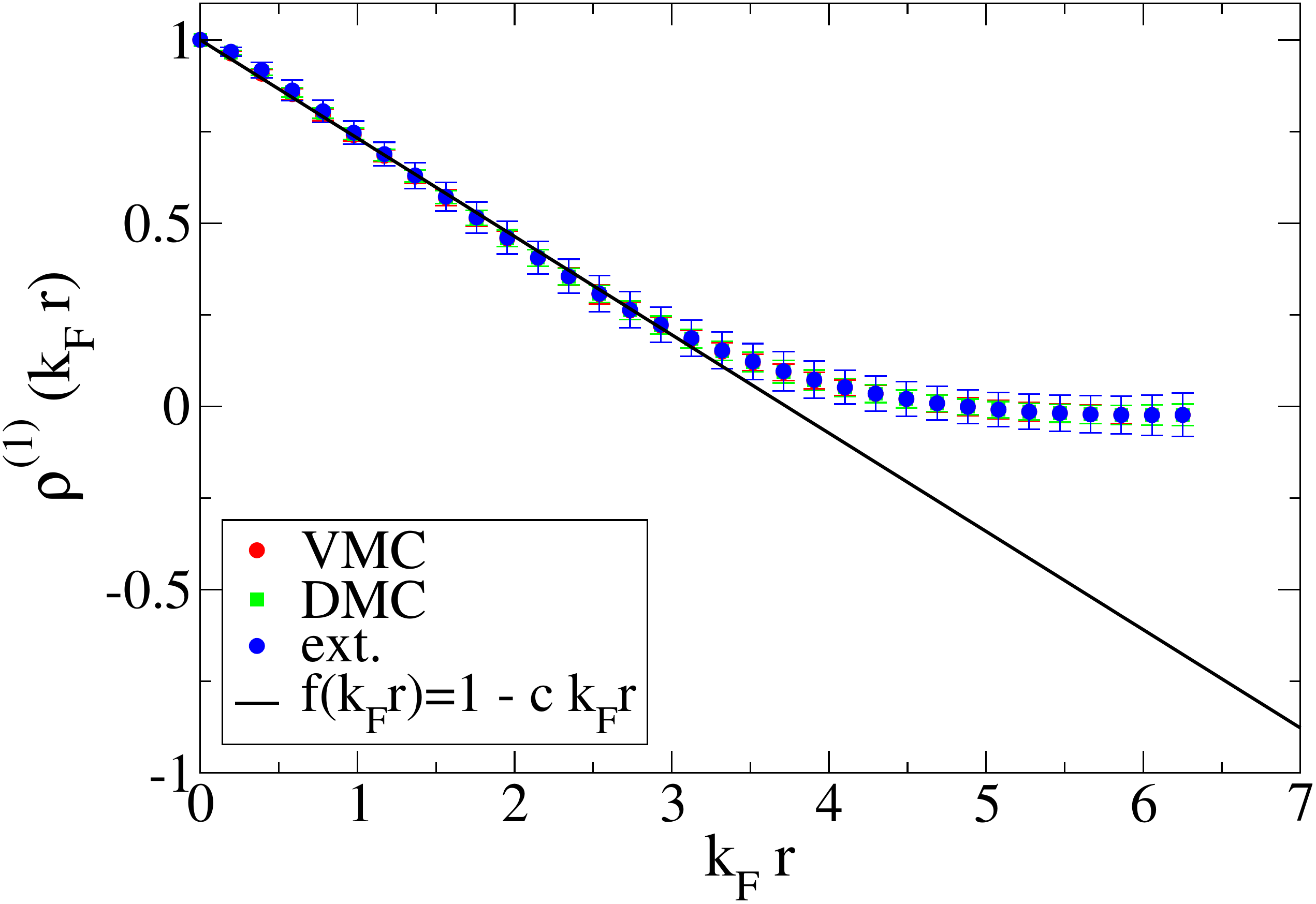}
\caption{
(color online) The radial one-body density matrix, symbols and $k_F\,r_e\approx 0.08$ as in Fig. \ref{fig:gofr}.
A line showing the linear fit with $c=0.2685$ is also shown, the dominant short-range
behavior is accurate up to approximately $k_F r \sim 3$.
}
\label{fig:rhor}
\end{figure}

The calculated radial one-body density matrix $\rho^{(1)}(k_F r)$
is shown in Fig. \ref{fig:rhor} using VMC and the mixed DMC results.
Again the results are nearly identical, with strikingly
linear behavior over a large range of small $k_F r$ values.
The fit gives $\zeta  = 0.895(16)$ again in good agreement with the
equation of state result.
\begin{figure}[ht]
\includegraphics[width=0.4\textwidth]{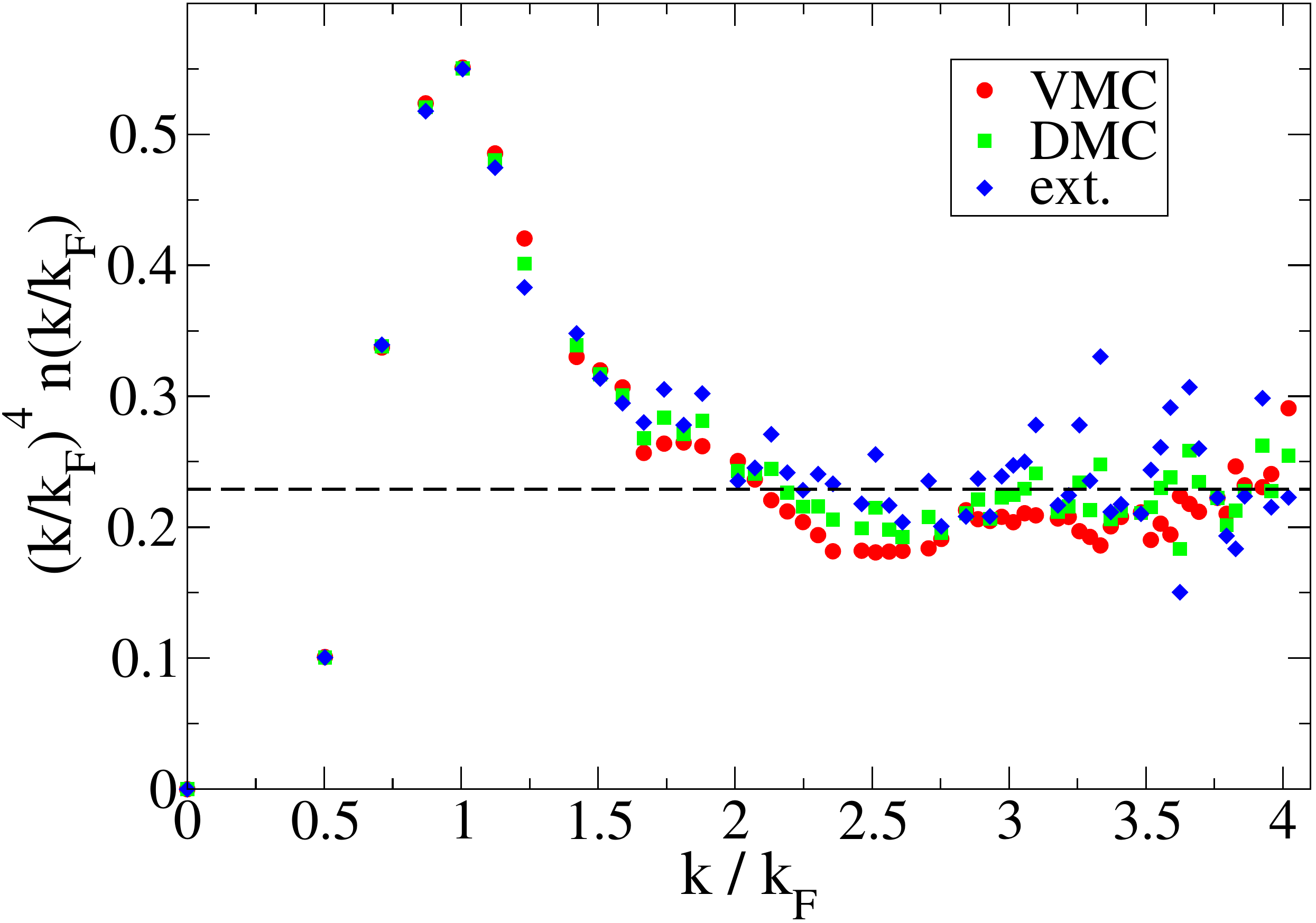}
\caption{(color online) The calculated momentum distribution summed over both spins
multiplied by $k^4/k_F^4$ showing the $k^{-4}$ tail.
Dashed line show $2C/k_F^4$ of Eq. \ref{eq:c-eos}
}
\label{fig:nofk}
\end{figure}
The calculated momentum distribution is shown in Fig. \ref{fig:nofk}.
The momentum distribution and the one-body density matrix are each
other's Fourier transform. The only difference in our calculations are
that the angular average has been done in real space for the one-body
density matrix to give the radial one-body density matrix, while
the momentum distribution is calculated for the $k$ vectors that
correspond to the periodic simulation cell. The extraction of the
$k^4$ tail is rather noisy; using the radial one-body density matrix
gives a more accurate fit. From our results it appears that 
the contact term dominates the behavior for
$k \gtrsim 2 k_F$. Our asymptote is
consistent with the value $0.229(1)$ expected from $\zeta = 0.901(2)$
(dashed line in Fig. \ref{fig:nofk}).

Recent experiments have measured the contact parameter from the
equation of state~\cite{Navon:2010},
momentum distribution directly using ballistic expansion
and indirectly through the rf line
shape and photoemission spectroscopy~\cite{Stewart:2010},
and from the static structure factor~\cite{Kuhnle:2010}.
Navon et al.~\cite{Navon:2010} extracted a value of $\zeta = 0.93(5)$
from their equation of state measurements. Our best value of $\zeta=0.901(3)$
is well within their experimental errors. Kunhle et al.~\cite{Kuhnle:2010}
calculate a slope of $S(k)$ versus $k_F/k$ at large $k$ for
$1/(k_Fa) = 0$ of $0.75(3)$ at $T=0.10(2)T_F$, giving a value of $\zeta = 0.80(3)$, while
Stewart et al. give values somewhat away from unitarity which also
give $\zeta$ lower than our value.

In conclusion, we have used Quantum Monte Carlo techniques to study 
the short-range correlations of unitary Fermi gases
as encoded in Tan's contact parameter.  The extractions from 
various observables  all give the same result within statistical errors.
These Monte Carlo methods give particularly low variance values for
the energy of the system and with minimal bias.
Therefore extracting the contact parameter
from the equation of state is the simplest and most reliable. However,
we have shown that its value extracted from the two-body radial distribution
function, the one-body radial density matrix, and the momentum distribution
also give the same results albeit with somewhat larger error bars.
For each of these quantities we have also determined the regime over which
the leading contact behavior is dominant, which should be useful to future
experiment in extracting the contact behavior and leading corrections.

{\it Acknowledgements:}
We thank J. E. Drut for valuable discussions.
This work is supported by the U.S. Department of Energy,
Office of Nuclear Physics, under contracts DE-FC02-07ER41457 
(UNEDF SciDAC), and DE-AC52-06NA25396 and by the National Science
Foundation grant PHY-0757703. KES thanks the Los Alamos National Laboratory
and the New Mexico Consortium for their hospitality. Computer time was made
available by Los Alamos Open Supercomputing.

\bibliographystyle{apsrev4-1}

%

\end{document}